# On the Statistical Settings of Generation and Load in a Synthetic Grid Modeling

Seyyed Hamid Elyas, Zhifang Wang[*]
Electrical and Computer Engineering
Virginia Commonwealth University, Richmond, VA, USA
Email: {elyassh, zfwang}@vcu.edu

Robert J. Thomas
Electrical and Computer Engineering
Cornell University, Ithaca, NY, USA
Email: rjt1@cornell.edu

*Abstract-* **This paper investigates the problem of generation and load settings in a synthetic power grid modeling of high-voltage transmission network, considering both electrical parameters and topology measures. Our previous study indicated that the relative location of generation and load buses in a realistic grid are not random but correlated. And an entropy based optimization approach has been proposed to determine a set of correlated siting for generation and load buses in a synthetic grid modeling. Using the exponential distribution of individual generation capacity or load settings in a grid, and the non-trivial correlation between the generation capacity or load setting and the nodal degree of a generation or load bus we develop an approach to generate a statistically correct random set of generation capacities and load settings, and then assign them to each generation or load bus in a grid.**

*Index terms-* **Synthetic Power Grid Modeling, Generation Setting, Load Setting, and Statistical Analysis.**

## I. Introduction

Synthetic power grid modeling, which are entirely fictitious but may have the same topology and electrical statistics of a realistic power grid, will help address the urgent need of grid data faced by many power system researchers and scientists, i.e., to provide sufficient case studies without disclosing any sensitive information of a realistic power grid.

To accomplish the goal of developing synthetic networks for the modeling of realistic power grid systems, extensive research has been conducted in order to recognize the salient grid related properties, to collect the statistics regarding the grid topologies and electrical parameters, hence to develop some useful models [1-18]. Reference [14] provides a comprehensive study on geographically approaches in synthetic power grid modeling. This paper describes several structural statistics and uses them to present a methodology to generate synthetic line topologies with realistic parameters. In [18] the authors propose a systematic methodology to augment the synthetic network base case for energy economic studies. In this paper the cost model of generators is determined based on the fuel type and generation capacity. This model can be utilized in electricity market and power system operation analysis.

As mentioned in [15], a valid synthetic grid model needs to include at least the following critical components: (a) the electrical grid topology which is fully defined by grid admittance matrix; (b) the generation and load settings which indicate their correlated siting and sizing; (c) the transmission constraints which include the capacities of both transmission lines and transformers and etc.

This paper presents our recent study results on the statistics of generation capacities and settings in a synthetic grid modeling. A set of approaches has been developed to generate a statistically correct random set of generation capacities and assign them to the generation buses in a synthetic grid according to the approximate scaling function of total generation capacity versus network size, the estimated exponential distribution of individual generation capacities, the non-trivial correlation between the generation capacity and the nodal degree of a generation bus. The proposed approaches may be readily applied to determining the load settings in a synthetic grid model and to studying the statistics of the flow distribution and to estimating the transmission constraint settings.

The main contributions of this paper are summarized as follows:(1) a set of statistical analysis on the generation capacities and its correlation with topology metrics is presented; (2) a statistical approach is proposed to determine the generation capacity at any generation bus in a synthetic grid modeling which takes into account both topology and electric measures; (3) a statistical-based approach is presented to determine the load at any load bus considering both topology and electric measures.

The rest of the paper is organized as follows: Section II presents the system modeling of a power grid; Section III provides the statistical analysis of generation capacities and their correlation with nodal degrees. This section also proposes an algorithm to generate and assign the generation capacities to generation buses; IV describes a statistical-based algorithm to determine the load size at any load bus. Finally, section V concludes the paper and discusses future works.

## II. System Modeling

The electrical topology of a power grid, with $N$ buses and $M$ branches which represent transmission lines and transformers in a high-voltage transmission network, is fully described by an admittance matrix $Y_{N \times N}$, which is defined as

$$Y = A^T \Lambda^{-1}(z_l) A, \qquad (1)$$

where A is the adjacency matrix. For a given power network with undirected and unitary links, it is defined as $A_{ij} = 1$ if nodes $i$ and $j$ are connected, 0 otherwise; $\Lambda^{-1}(.)$ denotes the diagonal inverse matrix with a specific vector and $z_l$ the vector of branch impedances in a grid. Therefore, the power flow distribution in a grid follows its network constraints as:

$$P(t) = B'(t)\theta(t) \qquad (2)$$

$$F(t) = \Lambda(y_l) A \theta(t) \qquad (3)$$

where $P(t) = [P_g(t), -PL_L(t), P_C]^T$ represents the vector of injected real power from generation, load and connection buses and obviously the power injection from connection buses equals zero, i.e., $P_C = 0$. $\theta(t)$ is the vector of phase angles, and $F(t)$ the vector of real-power delivered along the branches. Besides the network constraints, grid operation also needs to account for the constraints of generation capacity, load settings, and transmission capacity such as

$$P_g^{Min} \le P_g \le P_g^{Max} \qquad (4)$$

$$P_L^{Min} \le P_L \le P_L^{Max} \qquad (5)$$

$$F^{min} \le F \le F^{Max} \qquad (6)$$

From what is presented above it is clear that the dynamics of a power grid not only depend on the "electrical" topology but also the generation and load settings including their locations and sizing. The location setting of generations and loads is equivalent to the bus type assignment in power grid modeling which we have already address in [15] - [16]. In this paper we will expand our previous works on electrical setting of synthetic power grids to determine the setting of installed generators and loads in a given synthetic power grid.

## III. The Statistics of Generation setting

Generation settings in a synthetic grid modeling means to determine both the siting and the sizing of each generation unit. Our previous study [11] indicated that the relative location of generation and load buses in a realistic grid are not random but correlated. And an entropy based optimization approach has been proposed to determine a set of correlated siting for generation and load buses in a synthetic grid modeling. In this paper we focus on the problem to determine the sizing of generation units, i.e., the maximum capacity at each generation bus.

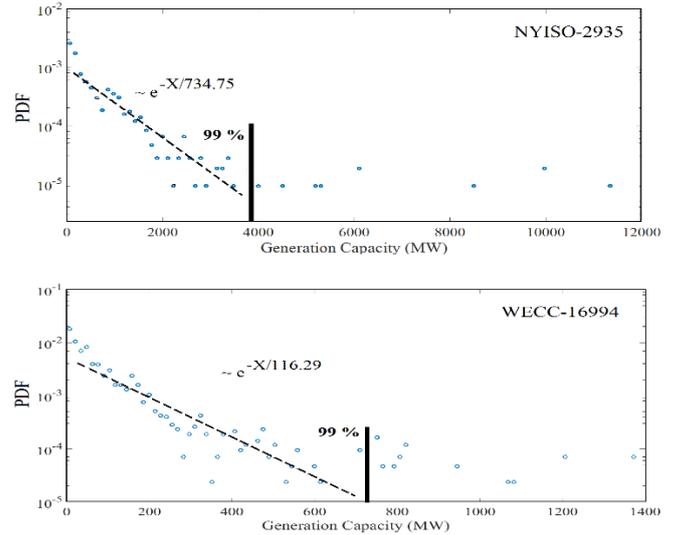

Fig. 1: Empirical PDF of generator capacities in NYISO-2935 and WECC-16994 bus system

In this section we first examine the statistical features of generation capacities in realistic power grids in terms of aggregate generation capacity, distribution of individual capacities, and their non-trivial correlation with nodal degrees.

In [17], we derived a scaling function of aggregate generation capacity in a grid versus its network size:

$$\log P_g^{tot}(N) = -0.21(\log N)^2 + 2.06(\log N) + 0.66 \qquad (7)$$

where $N$ is the network size, $P^{tot} = \sum_{n=1}^{N_g} P_{g_n}^{Max}$ denotes the total generation capacity, $N_g$ is the total number of generation buses, and the logarithm is with base 10. Equation (7) implies that the total generation capacity in a grid tends to grow as a power function when the network size is small. However, as the network size becomes larger, the scaling curve begins to bend down and grow slower than that power function.

Our study on the statistical distribution of generation capacity and demand within a power grid based on some realistic grid data such as the NYISO-2935, and the WECC-16994 systems, shows that more than 99% of the generation units (and the loads as well) follow an exponential distribution with about 1% having extremely large capacities (or demands) falling out of the normal range defined by the expected exponential distribution, as indicated by an empirical probability density function (PDF) of generation capacities shown in Fig. 1. A possible cause of these distribution exceptions may either come from an inherent heavy tailed distribution or result from boundary equalization in a network reduction modeling.

After studying the scaling property and distribution of generation capacities in a grid, it would be critical to examine the correlation between the generation capacities and other topology metrics. To simply the following statistical analysis and the algorithm development for generation capacity generation and assignment in a synthetic grid modeling, we define two normalized variables as

$$\overline{P_{g_n}^{Max}} = P_{g_n}^{Max} / \max_i P_{g_i}^{Max}, \tag{8}$$

$$\overline{k_n} = k_n / \max_i k_i. \tag{9}$$

So that both variables will assume values limited within [0, 1]. The statistics collected from the date of a number of realistic grids indicate that there exists a considerable correlation between the nodal degree of a generation bus and its capacity with a Pearson coefficient of $\rho\left(\overline{P_{g_n}^{Max}}, \overline{k_n}\right) \in$ [0.25, 0.5]. Fig. 2 shows the scatter plots of normalized generation capacity versus the normalized node degree of some sample grids as the NYISO-2935 and WECC-16994 systems, which exhibit similar distribution patterns. That is, most data points are densely located within the region of $\overline{P_{g_n}^{Max}} \in [0, 0.2]$ and $\overline{k_n} \in [0, 0.5]$, while very few located in the region of $\overline{P_{g_n}^{Max}} \geq 0.6$.

When two variable, say $\overline{P_{g_n}^{Max}}$ and $\overline{k_n}$ are considered, then we may put them together to get a pair of numbers, that is, a point $(\overline{P_{g_n}^{Max}}, \overline{k_n})$ in the two-dimensional space. These two-dimensional variables are considered mainly by their density function $f(\overline{P_{g_n}^{Max}}, \overline{k_n})$, which integrated on a set A gives the probability of the event that the value of $(\overline{P_{g_n}^{Max}}, \overline{k_n})$ is in the set A:

$$\Pr(A) = \Pr\left(\left(\overline{P_{g_n}^{Max}}, \overline{k_n}\right) \in A\right) \tag{10}$$

Fig. 3 illustrates the 2-D empirical probability density function (PDF) of normalized node degree versus normalized generation capacity in realistic power grids. Based on the obtained empirical PDF, a two- dimensional probability distribution table shown in Table I can be formulated to enable the algorithm development in next section to assign the generated capacity values to each generation bus in a grid according to its normalized nodal degree.

### III.A Assigning generation capacities to generation buses.

In this section we will introduce an approach to generate a statistically correct random set of generation capacities and assign them to the generation buses in a grid according to the approximate scaling function of total generation capacity versus network size, the estimated exponential distribution of individual generation capacities, and the non-trivial correlation between the generation capacity and the nodal degree of a generation bus.

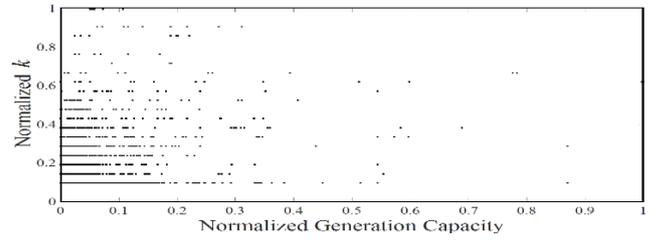

(a)

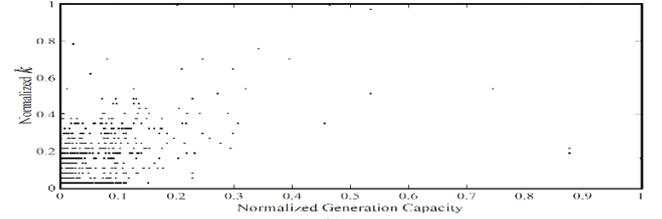

(b)

Fig. 2. Scatter plots of normalized node degree versus normalized generation capacities (a) WECC-16994 (b) NYISO-2935 bus system

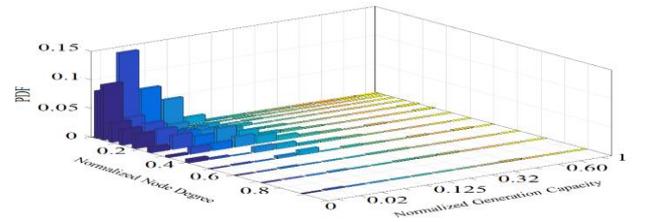

Fig. 3. 2-D empirical PDF of normalized node degree versus normalized generation capacity in WECC-16994 bus system.

Given a synthetic grid topology with $N$ buses among which $N_g$ buses have generation units, we may determine the aggregate generation capacity $P_g^{tot}(N)$ using equation (7) and generate a statistically correct random set of $N_g$ generation capacities which follows an exponential distribution of generation capacities with 1% of generated capacities switched to super large values. Then some scaling adjustment may be necessary to remain the same aggregate generation capacity given by $P^{tot}(N)$. Next an algorithm will be developed to assign the generation capacities to each generation bus with respect to the statistical pattern presented in Table I. Below is a step-by-step algorithm procedure description:

**Step 1**: Estimate the total generation capacity $P_g^{tot}$ using (7).

**Step 2**: Generate a statistically correct random set of generation capacities $[P_{g_n}^{Max}]_{1 \times N_g}$. It should be noted that 99 % of generated capacities follow the exponential distribution and remaining one percent is guaranteed to take supper large values (2~3 times greater than all generation capacities which follow the exponential distribution).

**Step 3**: Do the scaling of generated capacities if $\sum_{n=1}^{N_g} P_{g_n}^{Max} > 1.05 P^{tot}$ to make the aggregate generation capacity remain the range specified by $P_g^{tot}(N)$. And the scaling function is given as:

Table I. Probability analysis of normalized node degree and normalized generation capacity in WECC-16994 bus system

| | | $\overline{k_n}$ | | | | | | | | | | | | | Marginal Prob |
|---|---|---|---|---|---|---|---|---|---|---|---|---|---|---|---|
| | | 0.00–0.01 | 0.01–0.03 | 0.03–0.06 | 0.06–0.1 | 0.1–0.15 | 0.15–0.21 | 0.21–0.28 | 0.28–0.36 | 0.36–0.45 | 0.45–0.55 | 0.55–0.66 | 0.66–0.78 | 0.78–1.00 | |
| $\overline{P_{g_n}^{Max}}$ | 1.00–0.78 | 0.000 | 0.001 | 0.000 | 0.000 | 0.000 | 0.000 | 0.000 | 0.000 | 0.000 | 0.001 | 0.000 | 0.000 | 0.000 | 0.002 |
| | 0.78–0.66 | 0.000 | 0.000 | 0.000 | 0.001 | 0.000 | 0.000 | 0.001 | 0.001 | 0.001 | 0.000 | 0.000 | 0.000 | 0.000 | 0.004 |
| | 0.66–0.55 | 0.000 | 0.000 | 0.001 | 0.000 | 0.000 | 0.000 | 0.001 | 0.001 | 0.000 | 0.000 | 0.000 | 0.000 | 0.000 | 0.003 |
| | 0.55–0.45 | 0.000 | 0.001 | 0.000 | 0.004 | 0.008 | 0.000 | 0.002 | 0.001 | 0.000 | 0.001 | 0.000 | 0.001 | 0.000 | 0.018 |
| | 0.45–0.36 | 0.006 | 0.002 | 0.000 | 0.009 | 0.008 | 0.003 | 0.003 | 0.002 | 0.000 | 0.000 | 0.000 | 0.000 | 0.000 | 0.034 |
| | 0.36–0.28 | 0.003 | 0.011 | 0.012 | 0.017 | 0.013 | 0.007 | 0.003 | 0.002 | 0.000 | 0.001 | 0.000 | 0.000 | 0.000 | 0.072 |
| | 0.28–0.21 | 0.009 | 0.024 | 0.016 | 0.024 | 0.013 | 0.004 | 0.003 | 0.001 | 0.000 | 0.000 | 0.000 | 0.000 | 0.001 | 0.097 |
| | 0.21–0.15 | 0.025 | 0.027 | 0.016 | 0.013 | 0.009 | 0.002 | 0.002 | 0.000 | 0.000 | 0.000 | 0.000 | 0.000 | 0.001 | 0.097 |
| | 0.15–0.1 | 0.027 | 0.031 | 0.010 | 0.010 | 0.005 | 0.004 | 0.000 | 0.000 | 0.000 | 0.000 | 0.000 | 0.000 | 0.000 | 0.088 |
| | 0.1–0.06 | 0.033 | 0.017 | 0.003 | 0.003 | 0.005 | 0.000 | 0.001 | 0.000 | 0.000 | 0.000 | 0.000 | 0.000 | 0.000 | 0.063 |
| | 0.06–0.03 | 0.090 | 0.030 | 0.01 | 0.008 | 0.001 | 0.000 | 0.001 | 0.000 | 0.000 | 0.000 | 0.000 | 0.000 | 0.000 | 0.151 |
| | 0.03–0.01 | 0.082 | 0.140 | 0.070 | 0.04 | 0.010 | 0.001 | 0.000 | 0.000 | 0.000 | 0.000 | 0.000 | 0.000 | 0.000 | 0.360 |
| | 0.01–0.00 | 0.000 | 0.000 | 0.000 | 0.000 | 0.000 | 0.000 | 0.000 | 0.000 | 0.000 | 0.000 | 0.000 | 0.000 | 0.000 | 0.000 |
| Marginal Prob | | 0.283 | 0.291 | 0.147 | 0.141 | 0.077 | 0.022 | 0.017 | 0.008 | 0.001 | 0.003 | 0.000 | 0.001 | 0.002 | 1.000 |

$$[P_{g_n}^{Max}]'_{1 \times N_g} = [P_{g_n}^{Max}]_{1 \times N_g} \times \frac{P_g^{tot}}{\sum_{n=1}^{N_g} P_{g_n}^{Max}} \quad (11)$$

where $[P_{g_n}^{Max}]'_{1 \times N_g}$ is the updated generation capacities.

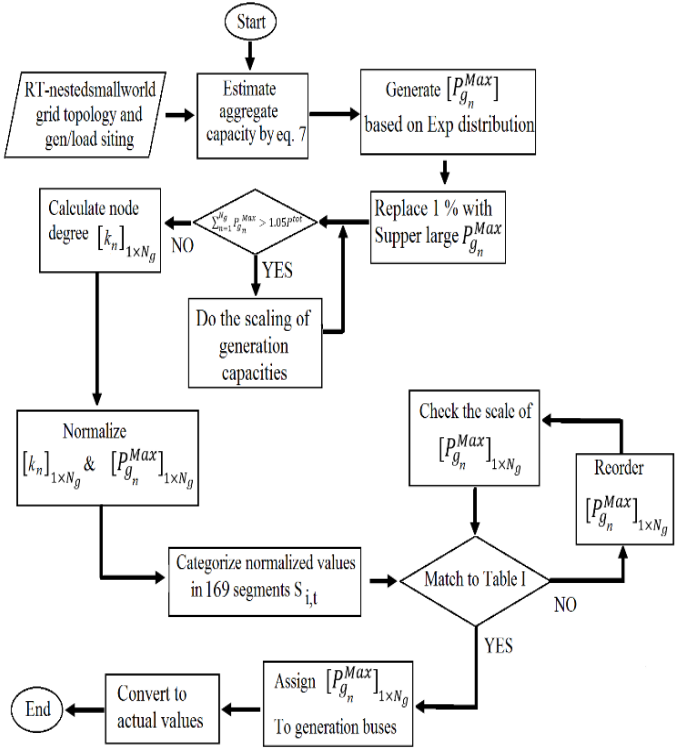

Fig. 4. Flowchart of the proposed algorithm to assign random generation capacities to generating buses

**Step 4**: Calculate the node degree for all generation buses 1 to $N_g$ based on the topology information of given synthetic power grid.

**Step 5**: Normalize both generation capacities and node degrees and categorize them evenly into 169 square regions with specific range of $\overline{P_{g_n}^{Max}}$ and $\overline{k_n}$.

**Step 6**: Check the similarity with Table I and reorder the mismatched segments.

**Step 7**: Assign the generated capacities to nominated generation buses with respect to their node degrees.

**Step 8**: Convert the normalized values to the actual values.

The flowchart of the proposed algorithm is depicted in Fig. 4.

## IV. The Statistics of Load setting

This section introduces a statistical –based approach to generate a set of static loads and assign them to the load buses. In this section we first investigate the statistical features of static loads in realistic power grids in terms of total demand, distribution of individual loads, and their non-trivial correlation with nodal degrees.

In [17] we reported the results of our analysis on scaling function of total demand in a grid versus its network size. Based on the obtained results the scaling of aggregate demand can be presented as a function of network size

$$\log P_L^{tot}(N) = -0.20(logN)^2 + 1.98(\log N) + 0.58 \quad (11)$$

where $P_L^{tot}(N) = \sum_{n=1}^{N_L} P_{L_n}$ denotes the total generation capacity and $N_L$ is the total number of load buses. The obtained results show that in realistic power networks the total demand tends to grow as a power function. It is important to point out that in a given grid topology although the total demand can be fully determined by the presented scaling function, failure to maintain a balance between total load and resources causes frequency to vary from its target value. Thus, it is crucial to consider both scaling function and aggregate sources to achieve a reasonable value for total demand.

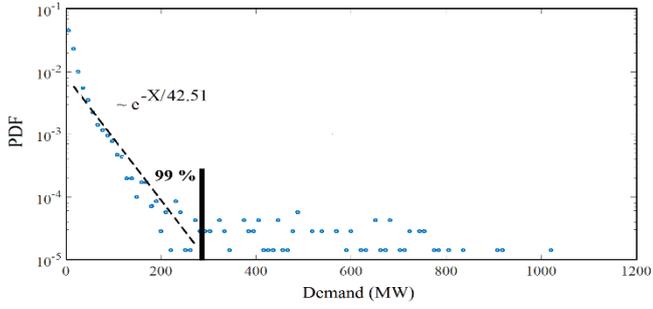

Fig. 5. Empirical PDF of loads in WECC-16994 bus system

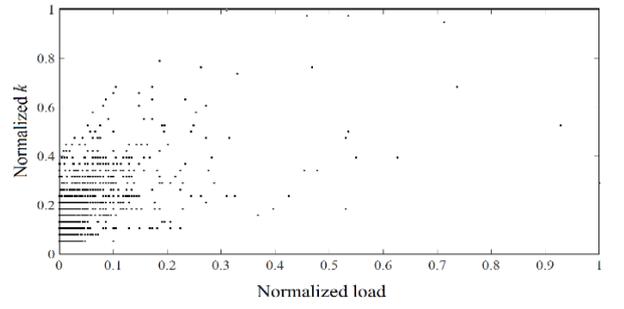

Fig. 7. Scatter plots of normalized node degree versus normalized load

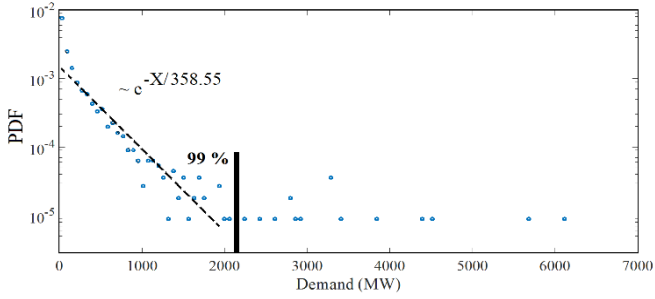

Fig. 6. Empirical PDF of loads in NYISO-2935 bus system

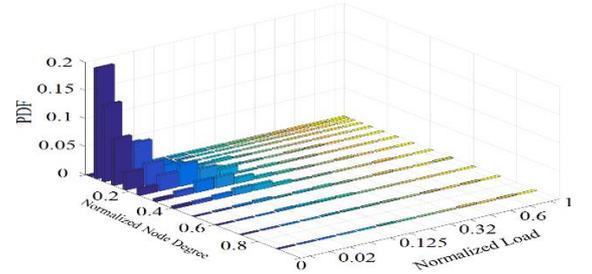

Fig. 8. The 2-D empirical PDF of normalized loads versus normalized node degrees in WECC-16994 bus system.

Table II. Probability analysis of normalized node degree and normalized load in WECC-16994 bus system

| | | $\overline{k_n}$ | | | | | | | | | | | | | Marginal Prob |
|---|---|---|---|---|---|---|---|---|---|---|---|---|---|---|---|
| | | 0.00–0.01 | 0.01–0.03 | 0.03–0.06 | 0.06–0.1 | 0.1–0.15 | 0.15–0.21 | 0.21–0.28 | 0.28–0.36 | 0.36–0.45 | 0.45–0.55 | 0.55–0.66 | 0.66–0.78 | 0.78–1.00 | |
| $\overline{P_L}$ | 1.00–0.78 | 0 | 0 | 0 | 0 | 0 | 0.001 | 0 | 0 | 0 | 0.001 | 0 | 0.001 | 0 | 0.002 |
| | 0.78–0.66 | 0 | 0 | 0 | 0 | 0.001 | 0.001 | 0.001 | 0.001 | 0 | 0.001 | 0 | 0.001 | 0 | 0.003 |
| | 0.66–0.55 | 0 | 0 | 0 | 0.002 | 0.002 | 0.002 | 0.002 | 0 | 0 | 0 | 0 | 0 | 0 | 0.008 |
| | 0.55–0.45 | 0 | 0.001 | 0.001 | 0.003 | 0.001 | 0.001 | 0.002 | 0.001 | 0 | 0.001 | 0 | 0 | 0.001 | 0.012 |
| | 0.45–0.36 | 0.003 | 0.004 | 0.006 | 0.010 | 0.010 | 0.003 | 0.001 | 0.001 | 0 | 0 | 0.001 | 0 | 0 | 0.041 |
| | 0.36–0.28 | 0.004 | 0.009 | 0.019 | 0.018 | 0.009 | 0.003 | 0.002 | 0.001 | 0 | 0.001 | 0 | 0 | 0 | 0.069 |
| | 0.28–0.21 | 0.012 | 0.027 | 0.037 | 0.022 | 0.013 | 0.001 | 0.002 | 0.001 | 0.001 | 0 | 0 | 0 | 0 | 0.118 |
| | 0.21–0.15 | 0.030 | 0.038 | 0.027 | 0.015 | 0.003 | 0.001 | 0.001 | 0 | 0.001 | 0.001 | 0 | 0 | 0 | 0.119 |
| | 0.15–0.1 | 0.082 | 0.066 | 0.022 | 0.004 | 0.002 | 0.003 | 0.001 | 0 | 0 | 0 | 0 | 0 | 0 | 0.183 |
| | 0.1–0.06 | 0.135 | 0.058 | 0.010 | 0.002 | 0 | 0 | 0 | 0 | 0 | 0 | 0 | 0 | 0 | 0.205 |
| | 0.06–0.03 | 0.196 | 0.033 | 0.005 | 0 | 0.001 | 0 | 0 | 0 | 0 | 0 | 0 | 0 | 0 | 0.235 |
| | 0.03–0.01 | 0 | 0 | 0 | 0 | 0 | 0 | 0 | 0 | 0 | 0 | 0 | 0 | 0 | 0.000 |
| | 0.01–0.00 | 0 | 0 | 0 | 0 | 0 | 0 | 0 | 0 | 0 | 0 | 0 | 0 | 0 | 0.000 |
| Marginal Prob | | 0.464 | 0.238 | 0.130 | 0.076 | 0.042 | 0.018 | 0.012 | 0.005 | 0.002 | 0.005 | 0.001 | 0.001 | 0.005 | 1.000 |

Our initial experiments on the statistical distribution of loads within realistic power grids show that, like generation capacities, about 99% of the loads follow an exponential distribution with about 1% having extremely large demands falling out of the normal range defined by the expected exponential distribution. Fig. 5 and 6 show the statistical distribution of loads in the WECC-16994 and NYISO-2935 bus system. The fitting curve is depicted as a dashed line for the distribution function of $P_L$. The straight line in the log plot implies that about 99% of loads in the WECC system tend to drop down as an exponential function with mean value of $\beta = 42.51$.

Given a realistic power grid with N buses among which $N_L$ buses have loads, we may examine the correlation between the total number of branches connecting a bus (that is, its node degree) and the total load attached to the bus, like what we have done in section III for generation capacities. To simply the following statistical analysis, we consider the normalized node degree presented in (9) and normalized load as:

$$\overline{P_{L_n}} = P_{L_n} / \max_i P_{L_i} \qquad (12)$$

So that the normalized loads will assume values limited within [0, 1]. Our statistical results show that in realistic power grids the Pearson's coefficient of correlation varies in the range of 0.3-0.6. Fig. 7 displays the scatter plot of normalized loads and normalized node degree which can be further used to generate the 2-D empirical PDF of some sample grids like WECC-16994 buses system.

By averaging the statistics of available realistic grid data, we may extract an empirical 2-dimensional probabilistic density function (PDF) for the normalized load values and nodal degree $\left(\overline{P_{L_n}}, \overline{k_n}\right)$. Based on the 2-D empirical PDF over the obtained uneven grid division (see Fig. 8) a two-dimensional probability distribution table shown in Table III can be formulated to enable an algorithm to assign the generated load values to each load bus in a grid according to its normalized nodal degree.

The approach to creating a statistically correct random set of loads and assigning them to the load buses begins by determining the aggregate load $P_L^{tot}(N)$ using equation (11) and generating a statistically set of $N_L$ loads which follows an exponential distribution of generation capacities with 1% of generated loads switched to super large values. In order to accurately formulate a synthetic power grid we need to assign generated loads to the load buses in a way consistent with that of a realistic grid. Therefore, the proposed approach will be developed to assign the generated loads to each load bus with respect to the statistical pattern presented in Table II. Expect for the statistical pattern, the procedure is exactly like that of generation capacities assignment in section III. The flowchart of the proposed algorithm is depicted in Fig. 9.

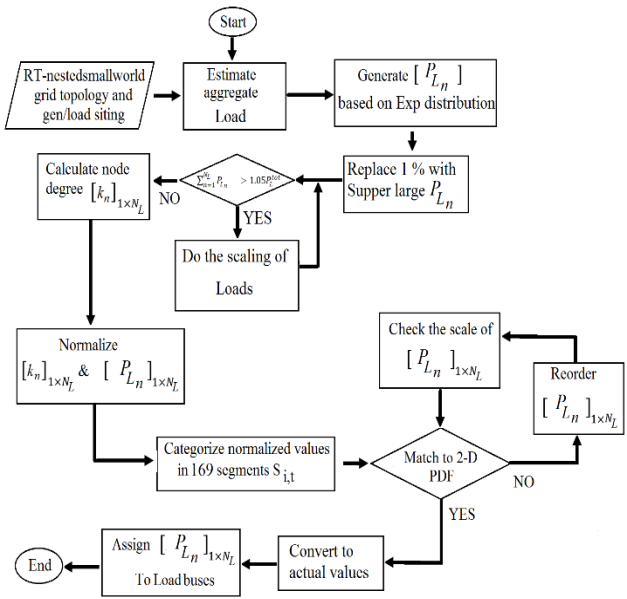

Fig. 9. Flowchart of the proposed algorithm to assign random loads to load buses

## V. Conclusion and future work

This paper presents our recent study results on the statistics of generation and load settings in synthetic grid modeling. In this paper we examine the statistical features of generation capacities and loads in realistic power grids in terms of aggregate generation capacity, distribution of individual capacities, and their non-trivial correlation with nodal degrees. Our study on the statistical distribution of both variables shows that more than 99% of the generation units/loads follow an exponential distribution with about 1% having extremely large capacities falling out of the normal range defined by the expected exponential distribution. Based on the obtained results presented in this paper there exists non-trivial correlations between the total number of branches connecting a bus (say node degree k) and the total generation/load attached to the bus.

Based on the above results, we develop two algorithm to generate a statistically correct random set of generation capacities and loads, and then assign them to each generation and load buses in a grid. In the future works these results will be used to determine the generation dispatch at each generation bus according to its generation capacity and the statistic of dispatch ratios, and to develop a statistical based algorithm to solve the transmission line capacity assignment problem in synthetic power grid modeling.